\documentclass[conference]{IEEEtran}

\IEEEoverridecommandlockouts
\usepackage{cite}
\usepackage{amsmath,amssymb,amsfonts}
\usepackage{algorithmic}
\usepackage{graphicx}
\usepackage{textcomp}
\usepackage{xcolor}
\usepackage{booktabs}
\usepackage{tabularx}

\def\BibTeX{{\rm B\kern-.05em{\sc i\kern-.025em b}\kern-.08em
    T\kern-.1667em\lower.7ex\hbox{E}\kern-.125emX}}
\begin{document}

\title{Beyond BFS: A Comparative Study of Rooted Spanning Tree Algorithms on GPUs}

\author{
\IEEEauthorblockN{Abhijeet Sahu, 
Srikar Vilas Donur}

\IEEEauthorblockA{Indian Institute of Technology Tirupati, India, 517619\\
\{cs22s501, cs23b049\}@iittp.ac.in}}

\maketitle

\begin{abstract}
Rooted spanning trees (RSTs) are a core primitive in parallel graph analytics, underpinning algorithms such as biconnected components and planarity testing. On GPUs, RST construction has traditionally relied on breadth-first search (BFS) due to its simplicity and work efficiency. However, BFS incurs an $O(D)$ step complexity, which severely limits parallelism on high-diameter and power-law graphs.

We present a comparative study of alternative RST construction strategies on modern GPUs. We introduce a GPU adaptation of the Path-Reversal RST (PR-RST) algorithm, optimizing its pointer-jumping and broadcast operations for modern GPU architecture. In addition, we evaluate an integrated approach that combines a state-of-the-art connectivity framework (GConn) with Eulerian tour–based rooting.

Across more than 10 real-world graphs, our results show that the GConn-based approach achieves up to $300\times$ speedup over optimized BFS on high-diameter graphs. These findings indicate that the $O(\log n)$ step complexity of connectivity-based methods can outweigh their structural overhead on modern hardware, motivating a rethinking of RST construction in GPU graph analytics.
\end{abstract}

\begin{IEEEkeywords}
spanning tree, rooted spanning tree, GPU, BFS, connected components
\end{IEEEkeywords}

\section{Introduction}
Rooted spanning trees (RST) are a foundational building block in parallel graph analytics, serving as the backbone for numerous algorithms shortest paths, ear decomposition, planarity testing and Bi-connected components \cite{pr-rst}. These operations are not merely theoretical; they are essential for uncovering the underlying topology of massive, real-world networks. 

The practical applications of RST construction are vast and cross-disciplinary. In computational biology, RSTs are used to analyze protein-protein interaction networks and reconstruct phylogenetic lineages. In power systems engineering, they facilitate the analysis of electrical grid resilience and contingency planning. Moreover, RST algorithms have been successfully applied to VLSI circuit design for clock-tree synthesis, robotics for path planning in configuration spaces, network telemetry for loop-free routing, and social network analysis to identify influential communities.

Despite their utility, the efficiency of RST construction on modern hardware is heavily dictated by graph structure. Formally, given an undirected connected graph $G =(V,E)$ and a designated root $r$, a rooted spanning tree (RST) is represented as a parent array $P$, where $P[r]=r$ and every other vertex has exactly one parent. Breadth-First Search (BFS) has long been the dominant approach for RST generation, and for good reason. The conceptual simplicity and easy to implement, makes it a default choice on GPU architecture. BFS guarantee of finding the shortest path (in terms of hops) from the root to all other vertices.

Yet, despite its ubiquity, BFS exhibits an inherent dependence on graph depth. BFS progresses via a level-synchronous traversal: all vertices at distance $k$ from the root must be discovered before any vertex at distance $k+1$ can be processed. While this structure is algorithmically clean, it introduces sequential dependencies that are fundamentally at odds with massively parallel hardware. On graphs with small diameters, this synchronization overhead is often negligible. However, on sparse real world graphs with large diameters, this can translate into thousands of serialized kernel launches and global synchronizations, significantly degrading performance.
At the opposite end of the design spectrum lie algorithms developed for the \textit{Connected Components (CC)} problem, which can be effectively adapted for this purpose. A crucial observation often overlooked in the literature is that connectivity itself is not the hard problem, rooting is. While BFS solves connectivity and rooting simultaneously, these tasks can be decoupled.

 In their seminal work, \textbf{Shiloach and Vishkin (1982)} \cite{sv-cc} demonstrated that, with minimal additional bookkeeping, any connectivity algorithm can simultaneously identify spanning tree edges during the component merging (union) phase. This method, however, only produces an unrooted forest. This gap was addressed when \textbf{Tarjan and Vishkin (1984)} \cite{tv-bcc} introduced the \textit{Eulerian Tour} technique, which provides an elegant method to assign roots to unrooted spanning trees in logarithmic depth. Together, the CC algorithm and Euler Tour technique form a powerful foundation for designing \textit{depth-oblivious} and \textit{work-efficient} RST algorithms with $O(\log n)$ parallel complexity.

However, while theoritically sound, another practical constraint, as highlighted by \textbf{Cong and Bader (2004)} \cite{pr-rst}, arises from the Euler Tour technique used for rooting. This approach requires transforming a standard adjacency list into a \textit{circular adjacency list}, where additional pointers define an Euler circuit that visits each directed edge exactly once. Constructing and maintaining such a structure incurs extra preprocessing overhead and irregular memory access patterns, both of which can influence performance on parallel architectures. To overcome these challenges, \textbf{Cong et al. (2004)} \cite{pr-rst}, building upon the Shiloach–Vishkin connectivity framework, proposed a fundamentally new algorithm called \textsc{pr-rst} for constructing Rooted Spanning Trees (RSTs) that eliminates the need for separate connectivity and rooting phases. Rather than first forming an unrooted spanning forest and subsequently rooting it using Euler tours, the \textit{Path-Reversal RST} algorithm treats both tasks as a single unified problem.

The table below summarizes the theoretical complexities of the primary approaches for RST construction.

\begin{table}[htbp]
\centering
\caption{Depth and work comparison of RST algorithms.}
\label{tab:rst_comparison}
\begin{tabular}{lcc}
\toprule
\textbf{Algorithm} & \textbf{Parallel Depth} & \textbf{Work Complexity} \\
\midrule
\textbf{BFS} \cite{merill-bfs} 
& $\Theta(\mathrm{diam}(G))$ 
& $O(V + E)$ \\

\textbf{CC + Euler Tour} 
& $O(\log V)$ 
& $O((V + E)\log V)$ \\

\textbf{PR-RST} \cite{pr-rst} 
& $O(\log V)$ 
& $O((V + E)\log V)$ \\
\bottomrule
\end{tabular}
\end{table}

\subsection{Our Contribution}

This paper makes the following contributions:

\begin{itemize}
  
  \item We present a GPU implementation of the PR-RST algorithm of Cong and Bader, adapted from its original multicore formulation, and evaluate its behavior under realistic GPU execution constraints.

  \item We show that RST construction can be naturally decomposed into a two-phase process consisting of connected components computation followed by a rooting phase using an Eulerian tour, both of which map efficiently to GPU architectures.

  \item Through an extensive experimental study on more than 30 real-world graph datasets, we identify the most effective RST construction strategy for GPUs.

\end{itemize}

\section{Related Work}
Harish and Narayanan (2007)~\cite{harish2007} were the first to implement BFS on the GPU using CUDA. Subsequent work by Merrill et al. (2012)~\cite{merill-bfs}, improved by using fine-grained prefix sums for load balancing and expanding frontiers at the edge level rather than the vertex level, they achieved an asymptotically optimal $O(V+E)$ work complexity. Subsequent GPU graph processing frameworks and libraries (e.g., Gunrock and Groute) further explored and optimized BFS. These efforts reinforced BFS as the most practical traversal primitive on GPUs, while simultaneously exposing its architectural limitations.

Despite these advancements, all traversal-based methods (BFS-based RST) remain fundamentally sensitive to the graph diameter (D). In ``sparse" graphs, the algorithm is forced into D sequential iterations. 

An alternative to traversal-based RST construction relies on parallel connectivity algorithms. The first large-scale adoption of connectivity algorithms on GPUs was demonstrated by Soman et al., who showed that Shiloach--Vishkin style pointer-jumping methods map naturally to GPU architectures. Since then, extensive progress has been made on GPU-based connected components, including work on label propagation, union--find variants, and hybrid approaches that balance work efficiency and convergence speed. Among these, GConn~\cite{gconn} represents the current state of the art. 




\section{Methodology}
In this section, we describe the algorithms studied in this work and their parallel GPU implementations.

\subsection{Breadth-First Search (BFS)}
Given a source vertex, BFS explores the graph in increasing order of distance, producing a level structure and an implicit spanning tree. At a conceptual level, BFS proceeds in rounds (or levels). In each round, all vertices discovered in the previous round (the \emph{frontier}) attempt to visit their undiscovered neighbors. Newly discovered vertices form the next frontier. This process continues until no undiscovered vertices remain reachable.
We adopt the edge-centric BFS implementation by Merrill et al.~\cite{merill-bfs} as our baseline.

\subsection{Connectivity-Based Algorithms}
The central idea is based on two alternating operations: \emph{linking} (also known as hooking) and \emph{compressing} (also known as shortcutting or pointer jumping).

\paragraph{Hooking (linking).}
Given an edge $(u,v)$, if $u$ and $v$ currently belong to different components, one component is linked to the other by making the root of one point to the root of the other. This operation is called a \emph{hook}. When a $hook$ succeeds, the corresponding edge $(u,v)$ is marked as a \emph{spanning edge}, since it contributes to the construction of a spanning forest.

Hooking can be performed in multiple ways (e.g., based on vertex IDs, random priorities, or ranks), but the goal is always the same: reduce the number of components while avoiding cycles.

\paragraph{Pointer jumping.}
After a round of hooking, the resulting structure is a forest. However, these trees may be deep and poorly shaped, forming long chains or star-like configurations. Such depth negatively impacts performance in subsequent rounds, as traversing long parent chains becomes expensive. To address this, connectivity algorithms employ pointer jumping, where each vertex $i$ updates its parent pointer as $(P[i] = P[P[i]])$, effectively halving the distance from every vertex to its root in each iteration. 

After $O(\log D)$ pointer-jumping steps, where $D$ is the diameter of the tree, all vertices in a component point directly to a single root. This \emph{flattening} dramatically reduces tree depth and prepares the structure for the next hooking phase. The alternating sequence of hooking and compression enables fast convergence and maps naturally to parallel architectures.

\subsection{Path-Reversal Rooted Spanning Tree (PR-RST)}
The \textbf{Path-Reversal RST (PR-RST)} algorithm, originally proposed as a multicore method by \textbf{Cong and Bader}~\cite{pr-rst}, has not previously been explored on GPUs to the best of our knowledge. In this work, we adapt PR-RST for modern GPUs, introducing a series of GPU-specific optimizations. 

As a connectivity-based algorithm, PR-RST follows the same hooking and pointer-jumping framework described earlier. The core idea is to efficiently update the tree after such component merges by processing all root–vertex pairs ($(u_i, r_i) \in R$) in parallel. For each pair, the algorithm identifies all vertices lying on the tree path from $u_i$ to $r_i$ simultaneously. Once these on-path vertices are determined, their parent–child relationships are reversed in parallel: for each edge where (u = \text{parent}[v]), the direction is flipped by setting (\text{parent}[u] = v).

To support efficient path identification, each vertex (u) maintains an array of size $O(\log n)$, referred to as its special ancestors. This array stores ancestors of (u) at distances that are powers of two, constructed via pointer jumping. Using these special ancestors, all vertices on the path from $u_i$ to $r_i$ can be identified in $O(\log n)$ parallel iterations, enabling scalable and GPU-friendly path reversal. 

Next we discuss on the optimizations done to make these operations efficient on GPUs. 

\noindent \textbf{Hooking.} During hooking, we alternate between \emph{max} and \emph{min} hooking, an optimization that empirically improves convergence and load balance. For each edge $(u,v)$, a single endpoint representative $\max(\texttt{rep}_u,\texttt{rep}_v)$ or $\min(\texttt{rep}_u,\texttt{rep}_v)$ is selected as the hooking target.

\noindent \textbf{Path Reversal.}
Rather than explicitly storing parent chains, we record pointer-jumping history during shortcutting using an `onPath` array. In the subsequent iteration, this history is used to reconstruct the reversed path by propagating markings over $(\lceil \log n \rceil)$ rounds. A second kernel then reverses parent pointers along the marked path, reorienting the tree from the old root toward the grafted node. This design avoids serial traversal, maintains coalesced memory access, and keeps path reversal fully data-parallel.

\noindent \textbf{Pointer Jumping.}
After each hooking phase, we perform pointer jumping to the compress trees. Final pointers converge to component roots and are used as representatives for the next iteration. Recording shortcut history is essential for enabling efficient path reversal without additional synchronization. 
To reduce kernel launch and synchronization overhead, we perform five pointer-jump steps per thread before a global synchronization. This strategy empirically outperforms synchronizing after every individual jump.

\subsection{Eulerian Tour}
While Polak et al.~\cite{euler-meets-cuda} previously adapted the Euler Tour algorithm for GPUs, their implementation relied on the legacy mgpu library and was restricted to connected trees. We focus on a careful re-engineering this implementation by using the modern \textbf{CUB (CUDA Unbound)} library and generalizes the algorithm to handle disconnected forests.

Given a forest with N vertices and N-1 undirected edges, we generate 2(N-1) directed edges and lexicographically sort them. This groups all outgoing edges of each vertex contiguously and induces a deterministic adjacency ordering. From the sorted permutation, we implicitly construct adjacency lists using three arrays: `first[v]' and `last[v]', denoting the range of outgoing edges for vertex (v), and `next[e]', linking consecutive edges with the same source.

For each directed edge $e$, we compute its Euler tour successor as
\[
\text{succ}(e) =
\begin{cases}
\text{next}(\text{rev}(e)), 
& \text{if } \text{next}(\text{rev}(e)) \neq -1, \\[4pt]
\text{first}(\text{from}(\text{rev}(e))), 
& \text{otherwise}.
\end{cases}
\]

Intuitively, after traversing $(e=(u \rightarrow v))$, the tour continues from (v) by advancing past the reverse edge $(\text{rev}(e)=(v \rightarrow u))$ in v’s adjacency list, wrapping around if necessary. This successor function induces one cyclic list per tree.

To generalize to disconnected forests, we break each cycle independently. For every root (r), we identify its last outgoing edge `last[r]', compute the corresponding reverse edge $((\text{last}[r] + E/2) \bmod E)$, and set its successor to (-1). This converts each Euler cycle into a linear list, yielding one independent traversal per tree. 

We then apply parallel list ranking to assigns each edge its position (rank) within its respective Euler tour. Since the cycles were broken into separate lists, each tree gets its own independent ranking.

Finally, the parent array is derived from edge ranks. For each edge $(e=(u \rightarrow v))$ and its reverse $(\text{rev}(e)=(v \rightarrow u))$, if $(\text{rank}[e] > \text{rank}[\text{rev}(e)])$, then `e' corresponds to the return traversal and `u' is the parent[v]; otherwise, v is the parent[u].

\section{Experimentations}
We evaluate all the algorithms on an NVIDIA L40s GPU using diverse variety of graphs. All codes were compiled with NVCC v12.9 (\texttt{-O3}, \texttt{--sm\_89}). Each dataset underwent one warm-up and five timed runs, reporting the median runtime. We first present an overall performance comparison across the three rooted spanning tree algorithms. Next, we analyze performance as a function of graph diameter, highlighting the contrasting behaviors of level-synchronous and connectivity-based approaches. Finally, we examine the structural properties of the resulting spanning trees and evaluate their impact on downstream algorithms.

\begin{table}[t]
\centering
\caption{Graph statistics. Diameter is approximated by the depth of the \textsc{bfs} spanning tree; M denotes millions.}
\begin{tabular}{lrrcl}
\hline
Dataset & \#Vertices & \#Edges & Diameter & Note \\
\hline
WB   & 0.69M & 13.30M & 973     & web-BerkStan \\
AS   & 1.70M & 22.19M & 757     & as-Skitter \\
HT   & 0.46M & 25.02M & 157     & higgs-twitter \\
CD & 0.54M & 30.49M & 14      & coPapersDBLP \\
SO   & 2.60M & 56.41M & 23{,}581 & sx-stackoverflow \\
RU   & 23.95M & 57.71M & 6{,}143 & road\_usa \\
LJ   & 4.85M & 85.71M & 1{,}877 & soc-LiveJournal1 \\
K20  & 1.05M & 89.75M & 253{,}378 & kron\_g500-logn20 \\
EU   & 50.91M & 108.11M & 19{,}932 & europe\_osm \\
K21  & 2.10M & 183.19M & 553{,}161 & kron\_g500-logn21 \\
CO   & 3.07M & 234.37M & 6       & com-Orkut \\
UK   & 18.52M & 523.65M & 38{,}360 & uk-2002 \\
\hline
\end{tabular}
\label{tab:dataset-stats}
\end{table}

\begin{figure}
    \centering
    \includegraphics[width=1\linewidth, scale=0.9]{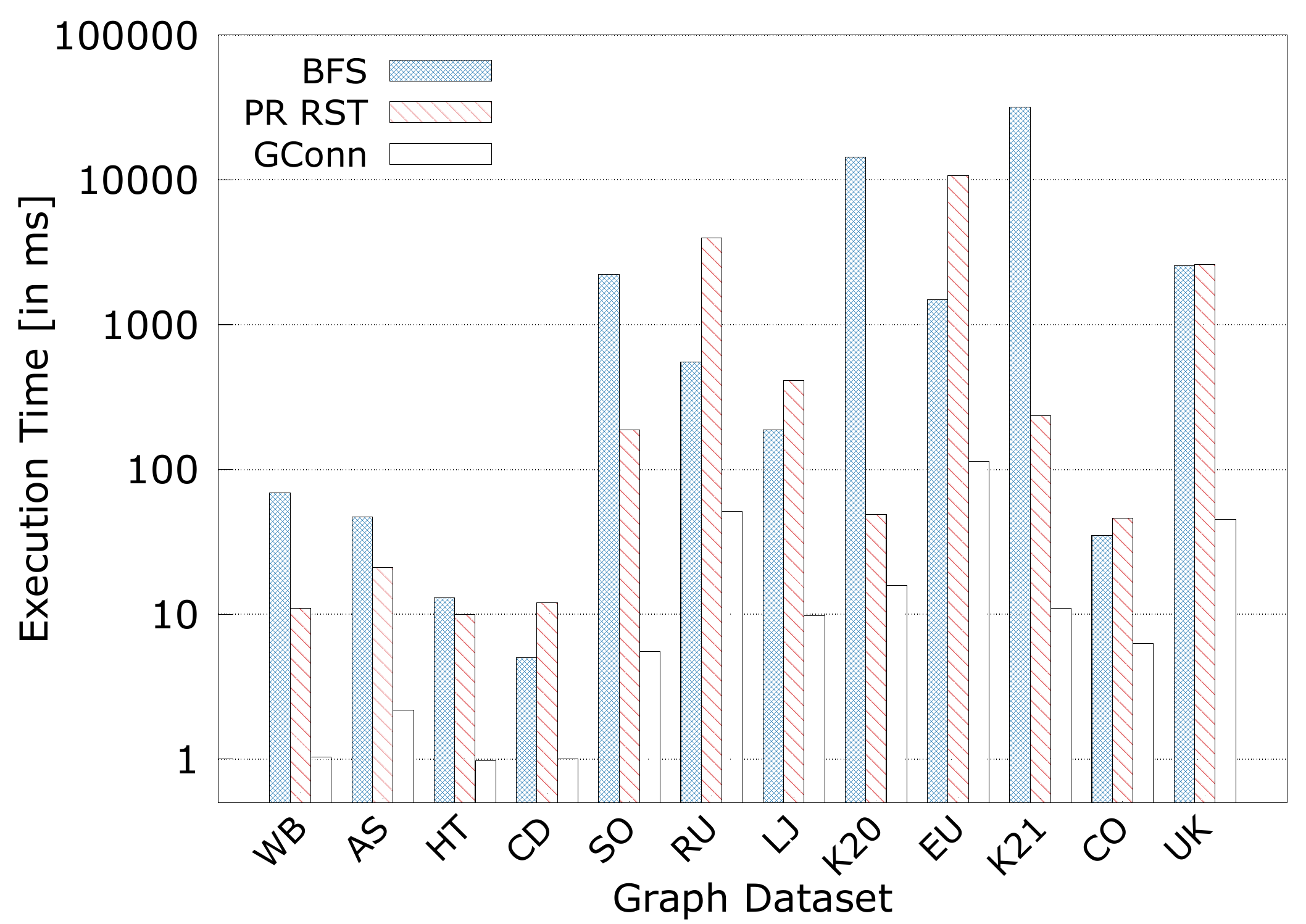}
    \caption{Running time comparison between \textsc{bfs}, \textsc{pr-rst}, and \textsc{gconn}.}
    \label{fig:comparison}
\end{figure}

\subsection{Performance Comparison of RST Algorithms}
Our results present a surprising outcome: \textsc{gconn} emerges as the fastest algorithm, despite the known theoretical overheads associated with Euler tour transformations. On average, \textsc{gconn} achieves a \textbf{300$\times$ speedup} over \textsc{bfs} and \textsc{pr-rst} as shown in \ref{fig:comparison}, outperforming even methods specifically designed to avoid these structural costs.

This finding strongly suggests that the perceived overhead of Eulerian tour transformations is a relic of older parallel models. On modern GPU architectures, this ``bottleneck'' is no longer a significant performance factor. This insight aligns perfectly with the recent work of Polak et al.~\cite{euler-meets-cuda}, who in ``Euler Meets GPU'' showed that these classical, theoretically-grounded techniques can be revived on GPUs to great practical effect. The elegance of the classical theory, once dismissed as impractical, is now unlocked by modern hardware. This challenges long-held assumptions and signals a paradigm shift for parallel spanning tree construction on GPUs.

\begin{figure}
    \centering
    \includegraphics[width=1\linewidth]{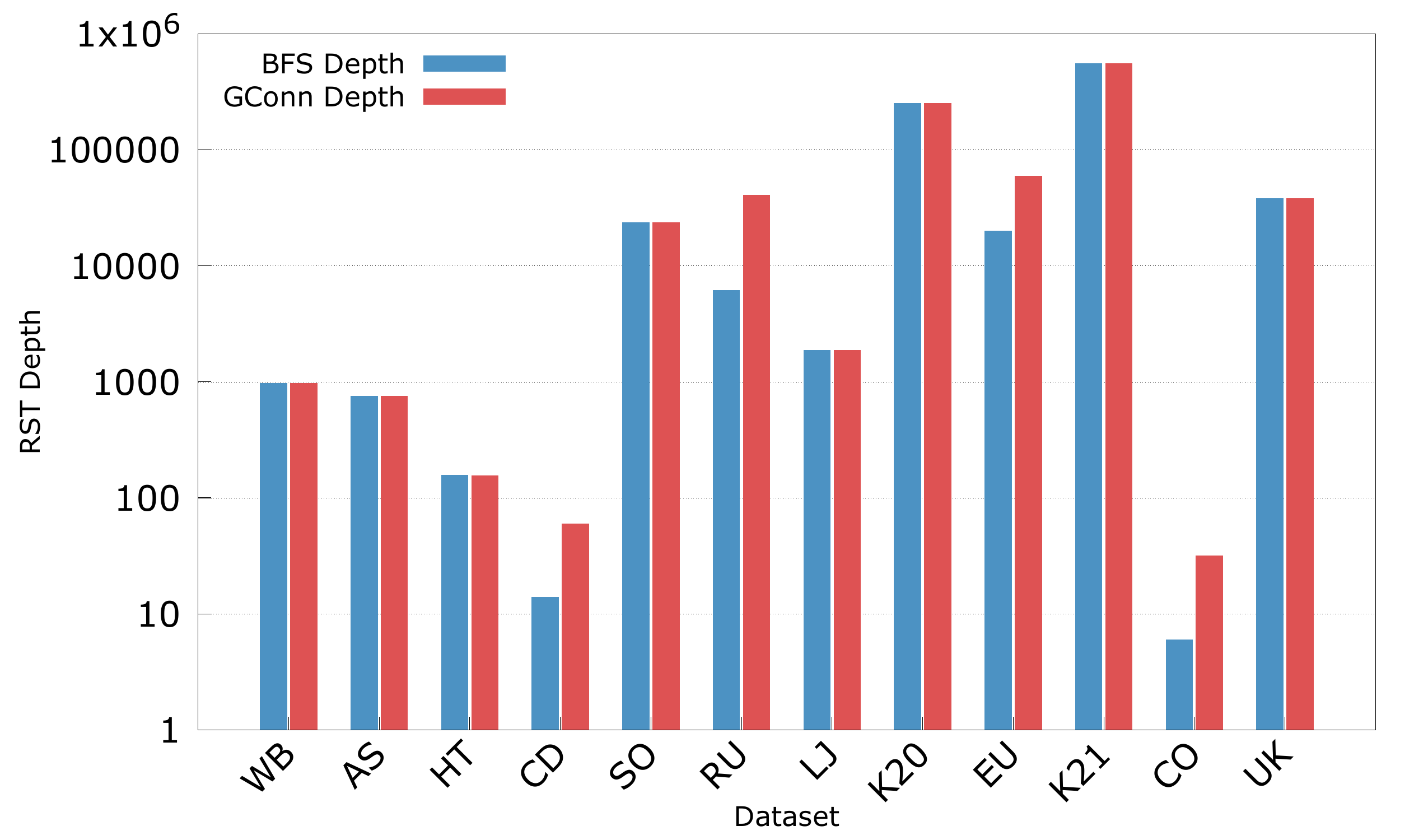}
    \caption{BFS RST depth vs GConn RST depth Comparison.}
    \label{fig:diameter}
\end{figure}

\subsection{Impact of Graph Diameter}
As shown in Fig.~\ref{fig:comparison}, the runtime of BFS increases with the depth of the resulting spanning tree. As discussed earlier, BFS operates as a level-synchronous traversal, exploring the graph one level at a time in a strictly ordered manner. While this approach is conceptually simple, it becomes a performance bottleneck on massively parallel architectures such as GPUs. Each level must be fully processed before the next can begin, introducing inherent sequential dependencies across levels.

For graphs with large diameters, such as road networks with over 1000 levels, this behaviour leads to thousands of sequential kernel launches. Consequently, the level-synchronous nature of BFS fundamentally limits its ability to exploit the massive parallelism available on modern GPUs when the graph diameter is large.

In contrast, the performance of \textbf{Gconn + Eulerian Tour} remains largely stable across varying graph diameters. As a connectivity-based approach, \texttt{gconn} avoids strict level-by-level synchronization and instead relies on parallel hooking and compression operations. This allows it to expose sufficient parallelism even for deep graphs, making it largely insensitive to increases in graph diameter. However, as shown in Fig.~\ref{fig:diameter}, this comes at the cost of producing deeper spanning trees, highlighting a depth–performance trade-off between traversal-based and connectivity-based approaches. 

\section{Discussion}
While BFS remains the dominant paradigm for a rooted spanning tree construction, our results suggest a necessary shift in perspective. Across multiple datasets, we observe that connectivity-based algorithms often produce significantly deeper trees. Although this structural difference raises new algorithmic questions for downstream tasks traditionally designed around BFS trees, it also highlights the need to rethink graph algorithms so they can more effectively exploit an arbitrary spanning tree.

\section{Conclusion}
While GPU-based connectivity algorithms have advanced significantly, BFS is far from obsolete; its inherent strengths in shortest-path discovery or cycle detection remains vital. However, for applications where an arbitrary spanning tree suffices, the performance gap between traditional traversals and modern connectivity-based approaches paired with an Euler Tour is too significant to ignore. For dynamic or incremental graphs, specialized approaches like PR-RST may offer a promising path toward maintaining rooted trees without the overhead of full reconstruction. We propose these findings as a guideline for future HPC graph frameworks: the choice of algorithm should be guided by the problem, not tradition. 

\bibliographystyle{IEEEtranS}
\bibliography{ref}

\end{document}